# Using the Econometric Models for Identification of Risk Factors for Albanian SMEs
# (Case study: SMEs of Gjirokastra region)

LORENC KOÇIU
Department of Economic Policies and Tourism
"Eqrem Çabej" University
Rruga "Studenti', Lagjia "18 Shtatori", Gjirokastra
ALBANIA

KLEDIAN KODRA
Managing Partner For Grant Thornton
ALBANIA
Agricultural University of Tirana
ALBANIA

*Abstract:* - Using the econometric models, this paper addresses the ability of Albanian Small and Medium-sized Enterprises (SMEs) to identify the risks they face. To write this paper, we studied SMEs operating in the Gjirokastra region. First, qualitative data gathered through a questionnaire was used. Next, the 5-level Likert scale was used to measure it. Finally, the data was processed through statistical software SPSS version 21, using the binary logistic regression model, which reveals the probability of occurrence of an event when all independent variables are included. Logistic regression is an integral part of a category of statistical models, which are called General Linear Models. Logistic regression is used to analyze problems in which one or more independent variables interfere, which influences the dichotomous dependent variable. In such cases, the latter is seen as the random variable and is dependent on them. To evaluate whether Albanian SMEs can identify risks, we analyzed the factors that SMEs perceive as directly affecting the risks they face. At the end of the paper, we conclude that Albanian SMEs can identify risk

## 1 Introduction

SMEs represent that part of business organizations that are faced with continuous changes in the market where they operate. Consequently, they are more exposed toward the probability of failure, financial difficulties, lack of liquidity, difficulties to take loans, unqualified employees, bankruptcy and many other challenges. According to the data generated from this study, and the literature reviews regarding SMEs, is indicated that the birth and death rate of SMEs is higher than other types of enterprises. Also, this is related with the fact that the legislation regarding SMEs activity, is much more flexible than the legislation that is related to large enterprises activity.

The market where SMEs operate is a market with a very fierce competition, with great problems and difficulties in entering new products, in their distribution, in finding new market segments from existing SMEs and those of new ones, which are newly trying to enter the market. This for many reasons, among which we can mention the size of these types of businesses, limited financial opportunities as well as restrictions on attracting a well-qualified staff compared to large business.

The risk faced by SMEs in their daily activities, in decision making, their implementation in practice, in staff absorption, as well as in many other issues is very different from the risk faced by big business. There is also a great diversity within the SME group itself, because the notion of SME includes within itself a diverse number of businesses such as legal form, number of employees, business figure, value of assets they have [1]. .

In Albania, according to Law no. 8957, dated 17.10.2002 "On Small and Medium Enterprises", as amended, gives the definition of what is called micro-enterprise, small enterprise and medium enterprise. Micro-enterprises are those enterprises, which employ up to 9 employees and their annual economic turnover does not exceed the amount of 10 million ALL[1]. Small enterprises are those enterprises which employ from 10 to 49 employees and have a total turnover or annual balance sheet of less than 50 million ALL. Medium enterprises are those enterprises which employ from 50 to 249 employees, have a business figure or total annual balance of up to 250 million ALL.

So SMEs are divided into:
(1) Micro enterprises
(2) Small Enterprises
(3) Medium Enterprises

The group of businesses included within the SME group faces problems with a completely specific nature of doing business, due to difficulties in securing sufficient resources such as financial, human, technological, operational and due to high fiscal workloads. In this context SMEs face a high level of risk, for which they should be able to identify the factors that cause this risk.

The Albanian literature on the identification of factors that affect the risks that SMEs face is rather scarce. Such scarcity must be taken into account by Albanian researchers. This paper is a minor contribution to fill this gap in the existing literature on the identification of factors that affect to the risk that SMEs in Albania face. However, it is impossible to address all such issues in one paper, so it is the duty of other researchers to give their contribution towards the literature on the risk that SMEs in Albania face.

Even though small and medium enterprises fall within a single group, that of SMEs, there are essential differences between these two groups. A major component among such essential differences is the risk they face, which varies for micro-businesses, small businesses and medium-sized ones. This fact prompted us to conduct this study concerning the risk that small and medium enterprises face.

This study also has its limitations, the absence of which would probably have contributed to even more positive or adjusted results from those obtained from the paper. These limitations have increased the level of difficulty in obtaining and processing data. But which are these limitations? We summarize them as follows:

➢ Qualitative data was obtained through questionnaires. In some cases, respondents were reluctant to answer all of the questionnaire questions. In others, the respondents were unable to answer the questionnaire arguing that the information required had to be supplied by a part-time employee who was not on their business.

➢ In some cases, the respondents did not have information about the concepts included in the questionnaire, or they often confused these concepts and therefore needed the help of the interviewer to clarify these concepts.

➢ Inaccuracies in the information provided both from official sources and through surveys, as a result of "*hiding*" data due to the relatively high rate of informality in our economy in general and the practices pursued by SMEs in particular.

The rest of the paper is structured as follows: Section 2 outlines the aim of the paper, where we raise two research questions and two hypotheses; Section 3 outlines a literature review of similar papers, highlighting their results and conclusions and a comparison of those; Section 4 outlines the methodology for data processing, selection of the observed population sample, and the codification of the variables studied; Section 5 outlines data analysis using logistic regression with the help of statistical software SPSS v21 and statistical testing of the hypotheses raised; Section 6 outlines conclusions and recommendations for researchers and politicians

## 2 Paper Objectives

The main objective of this paper is to identify the factors that influence into the risk that the Albanian SMEs face.

For the successful realisation of the main objective of this paper are raised two research questions:

**A)** Are the Albanian SMEs capable to identify the risk?
**B)** Are the Albanian SMEs capable to find out the main factors that cause the risk?

In focus of the above research questions and the general objective of this paper, the following hypothesis has been raised:

**H0:** The Albanian SMEs don't have real possibilities to identify the risk, which they face.

**H1:** The Albanian SMEs have real possibilities to identify the risk, which they face.

## 3 Literature Review

A brief and quick review of the literature on risk and its identification will be made on this issue. We will first try to give some definitions of risk, from the point of view of different researchers.

---

[1] ALL – Albanian money

[2] states that risk is injected into economic activity through various outflows of economic resources, which are carried out without knowing if they would be followed by positive inflows.

According to [3] the concept of risk can be seen as intertwined with uncertainty, giving the perception that it is uncertainty that leads to the emergence of risk. Events that lack predictability carry risks, although the results of these events can be predicted with an objective probability. Risk-affected outcomes have in themselves the potential to exhibit multiple values [4].

According to [5], the following aspects of risk should be considered:
- ❖ Risk can be determined both by an objective assessment (eg tossing a coin) and by a subjective assessment (eg: an individual assessment consisting of certain actions).
- ❖ Risk is defined at the individual and organizational level [6].
- ❖ Risk acceptance is influenced by group behavior compared to individual actions [7].

So different authors define risk in different ways, but everyone agrees with the idea that risk is related to the uncertainty and probability of the event occurring.

The risk identification process is the first step that an organization must take for a good management of its risk. But, regarding the process of identifying SME risk there are different opinions from different researchers.

According to [4] it is emphasized that entities should pay attention to the risk identification process, because risk can not be managed and controlled if it is not identified.

Furthermore states [8] risk identification provides the entity with an effective risk management process in relation to unknown sources causing unpredictable events and outcomes.

The main purpose of the risk identification process is not only the ability to identify losses caused by risk, ie negative risk, but also the ability to identify possible positive events, ie positive risk. The effect of not identifying positive risk is as valuable as the effect of not identifying negative risk [9]. The risk identification process means that all possible risks of the organization should be identified, as well as the opportunities offered to an organization.

The risk identification process is an integral part of the risk management process (RM). This process is very important and crucial for SMEs because as the complexity of products and services increases, so does the exposure of SME activity to risk [10].

SMEs, unlike large enterprises, suffer from numerous financial problems and a lack of human capital. It is because of these problems that SMEs face difficulties in applying or using risk management tools [11] (Brustbauer, 2016). Instruments used by large enterprises are usually inappropriate or impossible for SMEs to use because they can be very expensive or very complex [12] (Pereira et al, 2015).

SMEs are reluctant to and skeptical about implementing real risk management strategies to manage it [13] (Florio & Leoni, 2017), despite many studies showing that one of the main reasons for the failure and bankruptcy of SMEs is poor risk management, lack of planning of the risk identification and assessment process [14] - [15] (Sipa 2018) , Wasiluk 2017).

Risk identification and management practices should be simple and easily adapted to SMEs' operational plans to improve the performance of their business [16] (He & Lu, 2018). SMEs have understood this thing, and they have increased their awareness of risk identification and management, considering this process as very important in their activity [14] (Sipa 2018).

According to [17] potential sources of risk are considered to include;

a) **Social environment**. In the social environment according to these authors will be included the process of employee motivation, employee qualification, recruitment and selection, setting norms and culture of the organization in general. The social environment is identified as a source of risk because in each of the above mentioned links there is a possibility that misunderstandings, errors arise and as a result a potential risk appears.

b) **Physical environment**. The physical environment includes all the physical infrastructure of an organization such as the physical space of the offices, the location of the machinery, the age of the machinery and equipment, the humidity in the workplace, the ventilation system, the geographical position of the business location. Each of the above elements is a potential source of risk. If the used machinery is old and physically depreciated, this fact definitely poses a potential risk to the poor and timely production of goods or material goods.

c) **Political environment.** SMEs are heavily influenced by the country's political environment. As an element of the political environment we can mention the influence of fiscal or monetary policy. Both of these types of policies have a great impact and importance in the continuation of the activity of Albanian SMEs. So, Albanian SMEs should pay great attention to these policies, because otherwise it would lead them to bankruptcy and legal confrontation with the state.

d) **Legal environment**. The legal environment is another potential source of risk. The activity of SMEs in general, and the Albanians in particular, is regulated by a legal framework. Within this legal framework there are also various legal spaces where their use or non-use may constitute a potential source of risk for this business group.

e) **Operational environment**. The operational environment of SMEs includes elements related to their manufacturing or operating activity. As such can be mentioned the cases when the productive activity of SMEs can harm the health of its employees. This is a risk that needs to be given considerable attention because not only will it have financial consequences, but it can lead to opposition to the activity by the community concerned and even to its bankruptcy.

f) **Economic environment**. The economic environment includes elements such as the unemployment rate, inflation, etc. SMEs must have clear knowledge, complete and accurate information about these economic indicators of the country. Each of these indicators is a potential source of risk for Albanian SMEs [1].

## 4 Methodology

To realize successfully this paper through a structure questionnaire, are interviewd about 154 enterprises that belong to the SMEs classification, in order to collect primary data. The exact determination of the sample size is done based on the statistical formula, at the same time defining an error interval between 5% to 10%.

$$n = \frac{N}{(1 + Ne^2)}$$

n - stands for the sample size studied
N - stands for the full size of the population from which the sample was be selected
e - stands for the margin of error

If we would like to have a 95% confidence level in this study, we should have surveyed:

1. $n = \frac{N}{(1+Ne^2)} = \frac{2814}{(1+2814 x 0.05^2)} = 350 \; businesses$

But, this large number of businesses was impossible for us to survey, as a result, we had to reduce the confidence level to 92.15% and at the same time increase the margin of error from 5% to 7.85%. With this margin of error of 7.85% from the recalculation of the statistical formula, it results that:

2. $n = \frac{N}{(1+Ne^2)} = \frac{2814}{(1+2814 x 0.0785^2)} = 154 \; businesses$

The number of businesses in the Gjirokastra region included in the survey, acknowledging this level of error, was 154. To have a fair distribution, this selected sample was divided according to the specific weight that each business group had relative to the total population, table 1.

**Table 1. Dividing the random sample by activity**

| Total number of SMEs | Manufacturing | Construction | Commerce | Services |
|---|---|---|---|---|
| 154 | 28 | 6 | 62 | 58 |
| 100% | 18.23% | 4.19% | 40.37% | 37.21% |

*Source: Authors, 2020*

Through this questionnaire are collected qualitative primary data of dichotomous and ordinal type.

We have used the logistic regression to process these qualitative data in a statistical way. The logistic regression is used to analyze problems in which are more than one independent variables that influence to the dependent variable of dichotomous type, in case when this type of variable is considered as a casual variable that is dependent by them.

The logistic regression aims to find the correct model and to present it as the more appropriate linear mathematical equation that describes the relationship between the dependent variable and to the casual independent variables [1].

The equation of logistic regression is written like: $\ln(p/(1-p)) = B_0 + B_1X_1 + B_2X_2 + \ldots + B_nX_n$ and shows the probability of occurring an event under all independent variables. One of the most important elements to the logistic regression model is "odds ratio" or with more simple words "chances ratio".

Also, in this paper are collected secondary data through the updated literature for a theoretical supporting, through the official data provided by albanian public institutions and from other sources.

In Gjirokastra region operate about 2814[2] economic entities that belong to the SMSs group. These economic entities operate in different sectors like wholesale trading, retail trading, food and industrial items production, coffee-bar, hotels, tourism, etc. The study to this paper is done analyzing a large number of economic entities in the Gjirokastra region, which belong to micro, small and medium enterprises.

---

[2] INSTAT "The book of economic enterprises 2019", Tiranë 2019

The qualitative data are processed with SPSS statistical software version 21, which in its structure includes the processing of dichotomous and ordinal data through logistic regression. It is necessary to codify the variables for the SPSS statistical software, in order to process the data as much as correctly and simply. This codification includes the independent and dependent variables, as:

- ❖ P1 – Risk identification (dependent variable)
- ❖ P2 – The attempt to identify the risk factors when the organization has failed to an investment.
- ❖ P3 – Frequently changing law legislation.
- ❖ P4 – The increase of rate of loan interest.
- ❖ P5 – The exchange rate.
- ❖ P6 – The customers.
- ❖ P7 – The suppliers.

The P1 variable is the dependent variable and it belongs to the dichotomous type that takes only value 0 for the answer "NO", and 1 for "YES". While, other variables are considered as independent variables and they belong response to the ordinal type and are measured by 5 levels of Likert scale. The Likert scale is used to measure the qualitative data.

## 5 Data analysis

Table 1 shows the variables that are included into the logistic regression. As noted all the independent variables P2, P3, P4, P5, P6, which are included into this statistical model have a large statistical importance, because their alpha levels of significance have very low values (Sig = 0.000 < 0.05 or 5%). It means that these variables are referred to the 95% confidence level. While the P7 variable (the suppliers) is excluded by the model because of its unimportant statistical level that it has to predict the model, its alpha level of significance is too high (Sig>0.05), so the confidence level is less than 95%.

The variables that are included to this table are variables with a considerable influence that is determined by their logistic coefficients.

**Table 1. Variables in the Equation**

| Independent variables | | B | S.E. | Wald | df | Sig. | Exp(B) |
|---|---|---|---|---|---|---|---|
| Step 1[a] | P2 | -1.620 | .196 | 68.421 | 1 | .000 | .198*** |
| | P3 | 2.065 | .257 | 64.813 | 1 | .000 | 7.888*** |
| | P4 | 2.278 | .245 | 86.160 | 1 | .000 | 9.757*** |
| | P5 | 1.671 | .203 | 68.027 | 1 | .000 | 5.315*** |
| | P6 | 2.329 | .231 | 101.972 | 1 | .000 | 10.264*** |
| | Constant | -14.226 | 2.210 | 41.428 | 1 | .000 | .000 |

a. Variable(s) entered on step 1: P2, P3, P4, P5, P6.

Theses logistic coefficients show the probability that SMEs have in order to predict the chance to identify the factors that influence to the risk. These logistic coefficients are necessary to create the equation (1) of logistic regression:

$$\ln\frac{p}{1-p} = -1.620xP2 + 2.065xP3 + 2.278xP4 + 1.671xP5 + 2.329xP6 - 14.226$$

For testing the H1 hypothesis whether it is acceptable or not, we rely on the work of [18], according to which it is stated that when at least one of the coefficients near the independent variables which participate in the regression equation is different from 0 ($\neq 0$), then the hypothesis being tested is acceptable, or expressed differently, stands with an equal level of reliability with 95%.

Referring to equation (1) of the logistic regression it is observed that all the coefficients of this equation are different from zero. Also the value of the significance level is at zero level, which shows that statistically this forecasting model is significant within the 95% confidence interval. These mentioned above are summarized in Table 2.

**Table 2. Summary of statistical parameters of H1**

| Coefficients | Value | Sig | Statistical importance | H1 testing |
|---|---|---|---|---|
| B2 ≠ 0 | -1.620 | 0.000 | Sig<0.05 | Accepted |
| B3 ≠ 0 | 2.065 | 0.000 | Sig<0.05 | Accepted |
| B4 ≠ 0 | 2.278 | 0.000 | Sig<0.05 | Accepted |
| B5 ≠ 0 | 1.671 | 0.000 | Sig<0.05 | Accepted |
| B6 ≠ 0 | 2.329 | 0.000 | Sig<0.05 | Accepted |

*Source: Authors from SPSS statistical software v21 (2020)*

Also in the focus of hypothesis testing H1 will be analyzed the relationship that exists between the statistical parameters -2Log Likelihood and Chi-square (χ2), as well as their level of significance. For this purpose in the statistical program SPSS is done the complete test of their relationship, which is presented in table 3.

**Table 3. Iteration History**[a,b,c,d,e]

| Iteration | | -2 Log likelihood | Coefficients | | | | | |
|---|---|---|---|---|---|---|---|---|
| | | | Constant | P2 | P3 | P4 | P5 | P6 |
| Step 1 | 1 | 978.306 | -1.326 | -.316 | .267 | .365 | .249 | .494 |
| | 2 | 816.021 | -4.396 | -.760 | .694 | .873 | .600 | 1.073 |
| | 3 | 762.950 | -8.485 | -1.242 | 1.304 | 1.502 | 1.067 | 1.669 |
| | 4 | 749.365 | -12.339 | -1.525 | 1.828 | 2.031 | 1.476 | 2.128 |
| | 5 | 748.104 | -14.025 | -1.611 | 2.041 | 2.252 | 1.650 | 2.308 |
| | 6 | 748.089 | -14.223 | -1.620 | 2.065 | 2.278 | 1.670 | 2.328 |
| | 7 | 748.089 | -14.226 | -1.620 | 2.065 | 2.278 | 1.671 | 2.329 |
| | 8 | 748.089 | -14.226 | -1.620 | 2.065 | 2.278 | 1.671 | 2.329 |

a. Method: Enter
b. Constant is included in the model.
c. Initial -2 Log Likelihood: 970.478
d. Estimation terminated at iteration number 8 because parameter estimates changed by less than .001.

Starting the data processing with the value of -2Log Likelihood = 970.478 and completing the data processing to obtain the most statistically significant independent variables, with the value of -2Log Likelihood = 748.089, it turns out that the value of Chi- Sqaure will be as the difference of the value of -2Log Likelihood at the beginning of the model with the value of -2Log Likelihood in the last step of the model, ie 970.478 - 748.089 = 222.389, which is also evidenced in table 4 (Omnibus Tests). For this value of Chi- square results in a significance level (Sig = 0.000), which is compared to the default level α = 0.05 and the degrees of freedom (df=5), in order for the model to be significant within the 95% confidence interval.

**Table 4. Omnibus Tests of Model Coefficients**

| | | Chi-square | df | Sig. |
|---|---|---|---|---|
| Step 1 | Step | 222.389 | 5 | .000 |
| | Block | 222.389 | 5 | .000 |
| | Model | 222.389 | 5 | .000 |

*Source: Author from SPSS statistical software v21 (2020)*

Table 5 shows the distribution of Chi-Square according to the level of coefficient α and for the respective degree of freedom (df), which in this case belongs to the level, df = 5. From the comparison of the standard value of Chi-Square = 11.070 in Table 5, which belongs to the level α = 0.05 for the scale of df = 5 and the value of Chi-Square = 222.389 in table 4, which belongs to our regression model for the scale of df = 5, it turns out that 222.389 > 11.070. This means for the value 222.389 of Chi-sqaure, the significance level is less than the standard level α = 0.05, and in fact the significance level for this regression model is Sig = 0.000.

**Table 5. Chi Square distribution**

| | Probability level (α) | | | | | |
|---|---|---|---|---|---|---|
| Df | 0.5 | 0.10 | 0.05 | 0.02 | 0.01 | 0.001 |
| 1 | 0.455 | 2.706 | 3.841 | 5.412 | 6.635 | 10.827 |
| 2 | 1.386 | 4.605 | 5.991 | 7.824 | 9.210 | 13.815 |
| 3 | 2.366 | 6.251 | 7.815 | 9.837 | 11.345 | 16.268 |
| 4 | 3.357 | 7.779 | 9.488 | 11.668 | 13.277 | 18.465 |
| 5 | 4.351 | 9.236 | **11.070** | 13.388 | 15.086 | 20.517 |

Source: http://math.hws.edu/javamath/ryan/ChiSquare.html

So, we reject the null hypothesis (H0) in favour of the alternative hypothesis (H1). So, hypothesis H1: Albanian SMEs have a real opportunity to identify the risk they face, it is acceptable and statistically tested.

# 6 Conclusions and Recommendations

In summary, this paper highlighted that Albanian SMEs have real opportunities to identify risk. From equation (1) of logistic regression it became possible to verify and test the raised hypothesis. Carefully analyzing the independent variables included in equation (1) of logistic regression, their regression coefficients and odds ratio it turns out that;

- ❖ SMEs, which have had failures in previous investments have made and are making efforts to identify the factors that influenced these situations. Referring to the odds ratio of this variable, "P2", Exp (B) = 0.198 and its negative coefficient in the logistic regression equation (B = -1.620) it results that an increase in efforts to identify the factors leading to the failure of an investment will bring about a risk reduction and a facilitation in the application of the risk identification process.

- ❖ Frequent changes in legislation are a constant risk factor and SMEs do not neglect this fact, but are constantly wary of these changes. Referring to the odds ratio of this variable "P3", which corresponds to the identification of frequent change of legislation as a risk factor (Exp (B) = 7.888) and the positive logistic coefficient 2.065, it results that frequent change of legislation is a risk factor. with great impact, which means that when fiscal legislation changes then this change will affect SME risk about 7.9 times.

- ❖ Increasing the interest rate of the loan is a risk factor with a very large impact on the activity of SMEs and they must be careful when using the loan. Referring to the odds ratio of this variable,

"P4", Exp (B) = 9.757 and the positive regression coefficient 2.278, it results that an increase of one interest rate unit will affect the SME risk about 9.7 times.

❖ Exchange rate fluctuation is a risk factor for SMEs with a significant impact on their activity, mainly the exchange rate against the Euro currency. This is due to the fact that the vast majority of SMEs surveyed carry out sales-purchase relations abroad, due to their geographical position and proximity to neighboring Greece. Referring to the odds ratio of this variable (P5), Exp (B) = 5.315 and the positive regression coefficient 1.671, it results that an increase by one unit of the exchange rate will affect about 5.3 times the SME risk.

❖ Customer relationship is one of the most important factors in the risk of SMEs, therefore SMEs should pay close attention to how they will establish relationships with their customers, which has the greatest impact compared to with other predictive factors. Referring to the odds ratio of this variable (P6), Exp (B) = 10.264 and the positive logistic coefficient 2.329, it turns out that when the relationship with customers changes then this change will affect about 10.3 times the risk of SME.

At the end of this paper are highlighted some important recommendations:

• SME management should pay attention to risk identification and assessment as an integral part of the risk management process. SMEs should not focus only on one part of the risk management process but should conduct a thorough analysis of it.

• SME management must be vigilant against changes in fiscal legislation because frequent changes in legislation impact them adversely. Every type of business activity is regulated by a legal framework, which includes fiscal legislation. If SMEs fail to comply with legislation, then they face legal obligations, which is an additional high and consequently a very high-risk factor.

• SMEs should definitely prepare forecasts of the cash flows expected to be generated by future investments. Forecasting these flows reduces, to some extent, the risk that SMEs may face. This is of great importance today when liquidity problems are felt more than ever as the Albanian economy is still affected by the global crisis.

• SMEs should strive to develop a close working relationship with their customers because this greatly influences their future.

• The establishment and functioning of an SME association in Albania, which would function on a national and regional basis, would enable them to have a greater say in the preparation of the legal framework or fiscal package affecting this important sector of the economy of Albania.

• SMEs need to assess and manage risk also within the economic development strategies of the country and the region based on market analysis, prices, and demographic changes.

*References:*
[1] Koçiu, Lorenc. "Identifikimi dhe Vleresimi i Riskut te Ndermarrjeve te Vogla dhe te Mesme (Rast studimor: Rajoni i Gjirokastres)." *Doctoral thesis,* Tirane: Universiteti Bujqesor, 2015
[2] Kimball, Ralph C. "Failures in Risk Management" *New England Economic Review, January/February*, 2000: 1-12
[3] Smit, Yolande. "A Structured Approach to Risk Management for South African SMEs." Cape Town: Cape Peninsula University of Technology, March 2012
[4] Valsamakis, A. C, R. W Vivian, and G. S Du Toit. *Risk Management 2nd Edition.* Sandton :: Heinemann and Further Education: 31-32, 2000.
[5] Spekman, Robert E, and Edward W Davis. "Risky Business: Expanding the Discussio on Risk and teh Extended Enterprise." *International Journal of Physical Distribution & Logistics Management, Vol 34, Issue 5*, 2004: 414-433
[6] Spira, Laura F, and Michael Page. "Risk management: The reinvention of internal control and the changing role of internal audit." *Accounting, Auditing & Accountability Journal, Vol 16, No 4*, 2003: 640-661
[7] Giliberto, S.Michael, and Nkhil P. Varaiya. "The Winner's Curse and Bidder Competition in Acquisitions: Evidence from Failed Bank Auctions." *The journal of Finance, , Vol 12*, 1989: 637-648
[8] Williams Jr, Arthur C, Peter C Young, and Michael L Smith. *Risk Management and Insurance - 8th Edition.* New York: Irwin McGraw - Hill, 1998
[9] Dickson, Gordon C. A, and W. J Hastings. *Corporate Risk Management.* London: Witherby & Co, 1989
[10] Faisal, M. N. (2016). Assessment of supply chain risks susceptibility in SMEs using digraph and matrix methods. *International Journal of*

*Industrial and Systems Engineering*, 24(4), 441 – 468

[11] Brustbauer, J. (2016). Enterprise risk management in SMEs: Towards a structural model. *International Small Business Journal*, 34(1), 70-85.

[12] Pereira, L., et al. (2015). A risk diagnosing methodology web-based platform for micro, small and medium businesses: *Remarks and enhancements. Communications in Computer and Information Science*, 340-356.

[13] Florio, C., & Leoni, G. (2017). Enterprise risk management and firm performance: The Italian case. *The British Accounting Review*, 49(1), 56-74.

[14] Sipa, M. (2018). The factors determining the creativity of the human capital in the conditions of sustainable development. *European Journal of Sustainable Development*, 7(2), 1-13. Doi: 10.14207/ejsd.2018.v7n2p1.

[15] Wasiluk, A. (2017). Pro-innovative Prerequisites for Establishing the Cooperation between Companies (in the Perspective of Creation and Development of Clusters). In K. Halicka & L. Nazarko (ed.), Iwona Gorzeń-Mitka 349 © 2019 The Authors. Journal Compilation © 2019 European Center of Sustainable Development. 7th International Conference on Engineering, Project, and Production Management (T. 182, 755–762). Amsterdam: Elsevier Science Bv.

[16] He, C., & Lu, K. (2018). Risk management in SMEs with financial and non-financial indicators using business intelligence methods. *Management*, 16, 18.

[17] Tchankova, Lubka. "Risk identification – Basic stage in risk management." *Environmental Management and Health, Vol 13*, 2002: 290-297

[18] Bierens, J Herman. *The logit model: Estimation, Testing and Interpretation.* October 25, 2008. http://grizzly.la.psu.edu/~hbierens/ML_LOGIT.PDF

## Contribution of individual authors to the creation of a scientific article (ghostwriting policy)

**Author Contributions:**
Lorenc Koçiu is responsible for collecting the quality data, creating the database, processing the data with statistical software and organizing the paper.
Kledian Kodra is responsible for literature review, collecting the official data.